\documentclass[12pt]{article}
\pdfoutput=1

\usepackage{amssymb,amsmath,bm,bbm}
\usepackage{cite}
\usepackage{graphicx}
\usepackage{placeins}
\usepackage{longtable,color}

\definecolor{brown}{rgb}{0.6,0.7,0.}
\definecolor{gray}{rgb}{.38,.38,.38}

\def\simge{\mathrel{\rlap{\raise 0.511ex \hbox{$>$}}{\lower 0.511ex \hbox{$\sim$}}}}
\def\simle{\mathrel{\rlap{\raise 0.511ex \hbox{$<$}}{\lower 0.511ex \hbox{$\sim$}}}}
\def\slash#1{\setbox0=\hbox{$#1$}\dimen0=\wd0
      \setbox1=\hbox{/} \dimen1=\wd1 \ifdim\dimen0>\dimen1
      \rlap{\hbox to \dimen0{\hfil/\hfil}} #1                        \else
      \rlap{\hbox to \dimen1{\hfil$#1$\hfil}}
      /   \fi}

\newcommand{\lsim}{
\mathrel{\hbox{\rlap{\hbox{\lower4pt\hbox{$\sim$}}}\hbox{$<$}}}}

\newcommand{\gsim}{
\mathrel{\hbox{\rlap{\hbox{\lower4pt\hbox{$\sim$}}}\hbox{$>$}}}}

\def\eps{\varepsilon}
\def\epe{\varepsilon'/\varepsilon}

\newcommand{\gev}{\, {\rm GeV}}

\def\as{\alpha_s}

\allowdisplaybreaks[2]

\newcommand{\be}{\begin{equation}}
\newcommand{\ee}{\end{equation}}
\newcommand{\bea}{\begin{eqnarray}}
\newcommand{\eea}{\end{eqnarray}}

\newcommand{\bi}{\begin{itemize}}
\newcommand{\ei}{\end{itemize}}

\newcommand{\RE}{{\rm Re}}
\newcommand{\IM}{{\rm Im}}

\def\kpn{K^+\rightarrow\pi^+\nu\bar\nu}

\def\klpn{K_L\rightarrow\pi^0\nu\bar\nu}

\begin{document}
\begin{titlepage}
\vspace*{-1.0truecm}

{\Large \today}
\begin{flushright}
TUM-HEP-757/10\\
\end{flushright}

\vspace{0.4truecm}

\begin{center}
\boldmath

{\Large\textbf{The Impact of a 4th Generation on \\[0.3em]
Mixing and CP Violation in the Charm System}}

\unboldmath
\end{center}

\vspace{0.4truecm}

\begin{center}
{\bf Andrzej J.~Buras$^{a,b}$, Bj\"orn Duling$^a$, Thorsten Feldmann$^a$,\\
Tillmann Heidsieck$^a$, Christoph Promberger$^a$, Stefan Recksiegel$^a$
}
\vspace{0.4truecm}

{\footnotesize
 $^a${\sl Physik Department, Technische Universit\"at M\"unchen,
James-Franck-Stra{\ss}e, \\D-85748 Garching, Germany}\vspace{0.2truecm}

 $^b${\sl TUM Institute for Advanced Study, Technische Universit\"at M\"unchen,  Arcisstr.~21,\\ D-80333 M\"unchen, Germany}

}

\end{center}

\vspace{0.4cm}
\begin{abstract}
\noindent
We study $D^0-\bar D^0$ mixing in the presence of a fourth generation of quarks. 
In particular, we calculate the size of the allowed CP violation which is found at
the observable level well beyond anything possible with CKM dynamics.
We calculate the semileptonic asymmetry $a_\text{SL}(D)$ and the 
mixing induced CP asymmetry $\eta_f S_f(D)$ which are correlated with each other.
We also investigate the correlation of  $\eta_f S_f(D)$ with a number of prominent 
observables in other mesonic systems like $\epe$, ${\rm Br}(\klpn)$, 
${\rm Br}(\kpn)$,  ${\rm Br}(B_s\to \mu^+\mu^-)$, ${\rm Br}(B_d\to \mu^+\mu^-)$ 
and finally 
$S_{\psi\phi}(B_s)$ in the $B_s$ 
system. We identify a clear pattern of flavour and CP violation predicted 
by the SM4 model: While simultaneous large 4G effects in the $K$ and $D$ systems are
possible, accompanying large NP effects in the $B_{d}$ system are disfavoured. However this behaviour is not as pronounced as
found for the LHT and RSc models.
In contrast to this, sizeable CP violating effects in the $B_s$ system are possible unless extreme effects in 
$\eta_fS_f(D)$ are found, and ${\rm Br}(B_s\to\mu^+\mu^-)$ can be strongly enhanced regardless of the situation in the $D$ system.
{We find that, on the other hand, $S_{\psi\phi}(B_s) > 0.2$ combined with the measured $\epe$ significantly diminishes 4G effects within the $D$ system.}
\end{abstract}
\end{titlepage}

\section{Introduction}
The addition of a sequential fourth generation (4G)
of quarks and leptons to the SM (hereafter referred to as SM3) is
undoubtedly one of its simplest extensions. Reviews and summary statements of the thus extended model (hereafter referred to as SM4) can be found in 
\cite{Frampton:1999xi, Holdom:2009rf}.
While not addressing any of the known hierarchy and naturalness problems,
if present in nature a 4G is likely to have a number of
profound implications as clearly seen in the most recent literature.
In particular during the last years, a number of analyses were
published with the goal of investigating the impact of the existence of
a 4G on Higgs physics \cite{Kribs:2007nz,Hashimoto:2010at}, 
electroweak precision tests 
\cite{Alwall:2006bx,Kribs:2007nz,Chanowitz:2009mz,Novikov:2009kc,Erler:2010sk}, renormalisation group effects \cite{Hung:2009hy,Hung:2009ia} and 
flavour physics \cite{Babu:1987xe,Frampton:1999xi,Arhrib:2002md,Hou:2005yb,Hou:2006mx,Soni:2008bc,Herrera:2008yf,Bobrowski:2009ng,Eilam:2009hz,Soni:2010xh,Buras:2010pi,Hou:2010mm}.
Also detailed analyses of supersymmetry in the presence of a 4G
 have recently been performed in 
\cite{Murdock:2008rx,Godbole:2009sy}, where a good up to date collection of references to papers on the SM4 is contained in the second of these papers.

From the point of view of flavour changing neutral current (FCNC) processes, the
SM4 is particularly interesting {since, although it introduces non-Minimally 
Flavour Violating (MFV) interactions, it contains much fewer parameters than other 
New Physics (NP) scenarios like the Littlest Higgs model with T-parity (LHT), 
Randall-Sundrum (RS) models or the
general MSSM}. Moreover, the fact that the operator structure, similarly to 
the LHT model, is not modified with respect to the SM3, implies that the 
non-perturbative uncertainties in the SM4 are {at the same level as in the
SM3}. On the other hand, the absence of tree level contributions to FCNC processes,
in particular to particle-antiparticle mixing, and the absence of left-right
operators allows to satisfy {the experimental constraint from the CP violating
parameter $\varepsilon_K$ more easily than this} is possible in RS models. Finally, the 
non-decoupling effects of the 4G fermions and the fact that the masses of
new fermions are accessible easily to the LHC allows us to expect that
these NP scenario can be simultaneously tested by {high energy and flavour experiments}.

In \cite{Buras:2010pi}  we have performed a detailed analysis of non-MFV effects
in the $K$, $B_d$ and $B_s$ systems in the SM4, paying particular attention to correlations between flavour observables and addressing within this framework a number of anomalies present in the experimental data.
Similarly to the LHT model, the RS model with custodial symmetry (RSc) and
supersymmetric flavour models which have been analysed in \cite{Blanke:2006sb,Blanke:2006eb,Blanke:2009am,Blanke:2008zb,Blanke:2008yr,Altmannshofer:2009ne}, 
we have found that the SM4 can strongly violate correlations between various observables characteristic for models
with MFV and in particular with Constrained Minimal Flavour Violation (CMFV). At the same time it is still possible to satisfy all 
existing data on flavour violating processes and electroweak precision observables.

Probably the most striking signature of the SM4, compared to the LHT, RSc and
SUSY flavour models, is the possibility of having simultaneously sizeable 
NP effects in the $K$, $B_d$ and $B_s$ systems, even if truly spectacular 
effects are only possible in rare $K$ decays. The prominent exception
is $S_{\psi\phi}(B_s)$ which can also be enhanced by {more than} an order of 
magnitude --- {although the imposition of the $\epe$ constraint has significant 
impact on the size of this enhancement.}
This different pattern can be traced back to the fact that the mass scales involved in
the SM4 are generally significantly lower than in the LHT and in particular in the RSc.
Also the non-decoupling effects present in the SM4 which are absent in the LHT, RSc and SUSY play a role here.

The main goal of the present paper is the extension of our previous analysis \cite{Buras:2010pi} 
to $D^0-\bar D^0$ mixing and in particular to the mixing-induced CP violation
in the charm system, where CP violating effects are predicted to be tiny within the SM3\footnote{See however \cite{Bobrowski:2010xg}.}. 
The three basic questions that we want to address are as follows:
\begin{itemize}
\item
 Are large 4G effects in the charm system still possible {if consistence with all available constraints from tree level decays, particle-antiparticle mixing, $\epe$ and rare  decays in the $K$, $B_d$ and $B_s$ systems is required and also electroweak precision tests are taken into account?}
\item Can the large 4G effects in the $K$, $B_d$ and $B_s$ systems found by us in \cite{Buras:2010pi} still be maintained in the presence of constraints from charm data?
\item How are the 4G effects in the $K$, $B_d$ and $B_s$ systems correlated with the 
corresponding effects in the charm system? In fact, strong correlations
between the $D^0-\bar D^0$  and $K^0-\bar K^0$ systems in models with purely 
left-handed currents have been pointed out in \cite{Blum:2009sk} and
analysed also in \cite{Bigi:2009df}.
\end{itemize}
From the point of view of charm physics, the present paper follows the strategy of our previous papers on {this topic} in the context of the LHT model \cite{Bigi:2009df} and supersymmetric models \cite{Altmannshofer:2010ad}.
We will not repeat in detail the description of the SM4 and of the structure of the 4G mixing matrix since a detailed presentation of this model can be found in \cite{Buras:2010pi}, {where the same notation as in the present paper is used}. Similarly, we will not
repeat the formalism of $D^0-\bar D^0$ mixing as it is presented in \cite{Golowich:2007ka,Blum:2009sk,Bigi:2009df,Altmannshofer:2009ne,Amsler:2008zzb,Kagan:2009gb} in a notation suitable for our purposes.

Our paper is organised as follows. In Section \ref{sec:analytics} we briefly recapitulate the basic ingredients of the SM4 which are needed for our analysis and we present the effective Hamiltonian for $D^0-\bar D^0$ mixing in this model.
{Subsequently, we discuss in explicit terms the connection between $D$ and $K$ physics pointed out in \cite{Blum:2009sk}, which in the SM4 turns out to be not as transparent as in LHT \cite{Bigi:2009df}.}
To this end we consider different scenarios for the $V_{4G}$ mixing matrix which have been analysed in detail in our previous paper \cite{Buras:2010pi}.

Section \ref{sec:numerics} is devoted to our numerical analysis. Here we study a number
of correlations between different observables and we address the three
questions posed above. We summarise our results in Section \ref{sec:conclusion}.

\section{Relevant Formulae and Analytical Considerations}\label{sec:analytics}

\subsection{The Effective Hamiltonian}
The effective Hamiltonian for $\Delta C=2$ transitions following from the usual
box diagrams and including the contributions from the $b^\prime$-quark is given as follows
\begin{align}\label{eq:Heff}
\begin{split}
 {\mathcal H}_{\rm eff}^{\Delta C = 2}\:=&\:\frac{G_F^2}{16\pi^2}M_W^2\left[{\lambda_{s}^{(D)}}^2\eta^{(D)}_{ss}S_0(x_s)+{\lambda_{b}^{(D)}}^2\eta^{(D)}_{bb}S_0(x_b)+{\lambda_{b^\prime}^{(D)}}^2\eta^{(D)}_{b^\prime b^\prime}S_0(x_{b^\prime})\right.\\
+&\:{2\lambda_{b}^{(D)}}{\lambda_{s}^{(D)}}\eta^{(D)}_{bs}S_0(x_b,x_s)+{2\lambda_{b^\prime}^{(D)}}{\lambda_{s}^{(D)}}\eta^{(D)}_{b^\prime s}S_0(x_{b^\prime},x_s)\\
+&\:\left.{2\lambda_{b^\prime}^{(D)}}{\lambda_{b}^{(D)}}\eta^{(D)}_{b^\prime b}S_0(x_{b^\prime},x_b) \right]\left[\as^{(4)}(\mu)\right]^{-6/25} \left[1+\frac{\alpha_s^{(4)}(\mu)}{4 \pi} J_4 \right]\ Q(\Delta C = 2)\,+ h.c.\,,
\end{split}
\end{align}
where
\begin{align}
 Q(\Delta C = 2) &\,=\, (\bar uc)_{V-A}(\bar uc)_{V-A}\,,
\end{align}
and
\begin{align}
 {\lambda_{i}^{(D)}} &= V_{ci}^\ast V_{ui}\,,
\end{align}
which satisfy the unitarity relation
\begin{align}
 {\lambda_{d}^{(D)}}+{\lambda_{s}^{(D)}}+{\lambda_{b}^{(D)}}+{\lambda_{b^\prime}^{(D)}}&\,=\,0~.
\end{align}
For the QCD corrections we will use the approximate relations
\begin{align}\label{eq:QCDcorr}
 \eta^{(D)}_{b^\prime b^\prime} &\approx \eta^{(K)}_{tt}\,, & \eta^{(D)}_{b^\prime b}\approx \eta^{(D)}_{b^\prime s}&\approx \eta^{(K)}_{ct}\,, & 
\eta^{(D)}_{ss}\approx\eta^{(D)}_{bb}\approx\eta^{(D)}_{bs}&\approx\eta^{(K)}_{cc}\,,
\end{align}
which are found by inspecting the structure of QCD corrections in the
$D^0-\bar D^0$  and $K^0-\bar K^0$ systems. In this context let us remark that with $m_{b^\prime}$ defined as $m_{b^\prime}(m_{b^\prime})$ the QCD factors are only weakly dependent on $m_{b^\prime}$. Moreover we recall 
that the $\mu$-dependent QCD corrections in (\ref{eq:Heff}) are absorbed in 
the renormalisation group invariant parameter $\hat B_D$.
This parameter is defined by
\begin{equation}
\hat B_D=B_D(\mu)\left[\alpha_s^{(4)}(\mu)\right]^{-6/25}\left[1+\frac{\alpha_s^{(4)}(\mu)}{4\pi}J_4\right]\,.
\end{equation}
From lattice calculations \cite{Gupta:1996yt,Lellouch:2000tw,Lin:2006vc}, one has $B_D(\mu=2\gev)=0.845\pm0.024^{+0.024}_{-0.006}$, and with $J_4=6719/3750$ \cite{Buras:1990fn} and $\alpha_s(M_Z)=0.1184\pm0.0007$ \cite{Bethke:2009jm}, we find
\begin{equation}
\hat B_D=1.18^{+0.07}_{-0.05}\,,
\end{equation}
which will be used in our numerical calculations.

\subsection{CP Violation and Mixing Parameters}
The short distance (SD) contributions to the matrix element {of the effective Hamiltonian (\ref{eq:Heff})} can be written as
\begin{align}
 \langle \bar D^0 | {\mathcal H}_{\rm eff}^{\Delta C = 2} | D^0\rangle_{\rm SD} &\,\equiv\, \left| M_{12}^D\right| e^{2i\phi_D}\,=\, \left(M_{12}^D\right)^\ast\,,
\end{align}
where
\begin{align}
 M_{12}^D \,&=\, \frac{G_F^2}{12\pi^2}F_D^2 \hat B_D m_D M_W^2 \overline{M}_{12}^D\,,
\end{align}
with
\begin{align}\label{M12D}
\begin{split}
\overline M_{12}^D \,&=\,{\lambda_{s}^{(D)}}^{\ast 2}\eta^{(K)}_{cc}S_0(x_s)+{\lambda_{b}^{(D)}}^{\ast 2}\eta^{(K)}_{cc}S_0(x_b)+{\lambda_{b^\prime}^{(D)}}^{\ast 2}\eta^{(K)}_{tt}S_0(x_{b^\prime})\\
+&\,2{\lambda_{b}^{(D)}}^\ast {\lambda_{s}^{(D)}}^\ast \eta^{(K)}_{cc}S_0(x_b,x_s)+2{\lambda_{b^\prime}^{(D)}}^\ast {\lambda_{s}^{(D)}}^\ast \eta^{(K)}_{ct}S_0(x_{b^\prime},x_s)+2{\lambda_{b^\prime}^{(D)}}^\ast {\lambda_{b}^{(D)}}^\ast \eta^{(K)}_{ct}S_0(x_{b^\prime},x_b)\,.
\end{split}
\end{align}
Here we used the approximations given in (\ref{eq:QCDcorr}).
The full matrix elements then are given by
\begin{align}
\langle \bar D^0 | {\mathcal H}_{\rm eff}^{\Delta C = 2} | D^0\rangle &\,=\, \left( M_{12}^D+M_{12}^{\rm LD}\right)^\ast - \frac{i}{2}{\Gamma_{12}^{\rm LD}}^\ast\,,\\
\langle D^0 | {\mathcal H}_{\rm eff}^{\Delta C = 2} | \bar D^0\rangle &\,=\, \left( M_{12}^D+M_{12}^{\rm LD}\right) - \frac{i}{2}{\Gamma_{12}^{\rm LD}}\,.
\end{align}
Here $\Gamma_{12}^{\rm LD}$ 
and $M_{12}^{\rm LD}$ stand for long distance (LD) contributions with the former {arising exclusively from SM3 dynamics}. These contributions are very difficult to estimate and will be included in our phenomenological analysis using the strategy of \cite{Bigi:2009df,Altmannshofer:2010ad}.

\subsection{Connections Between D and K Physics and Beyond}
In \cite{Blum:2009sk}, the connection between $D^0-\bar D^0$ and $K^0-\bar K^0$ mixing has been discussed within the framework of approximately $SU(2)_L$-invariant NP models. Due to the {connection between up- and down-type quarks in the {SM3} through the CKM matrix}, the NP contributions to $D^0-\bar D^0$ and $K^0-\bar K^0$ mixing are not independent of each other. This observation has been used in \cite{Blum:2009sk} to derive lower bounds on the NP scale in various NP scenarios, which emerge if the experimental constraints on $D^0-\bar D^0$ and $K^0-\bar K^0$ mixing are applied to only the $(V-A)\otimes (V-A)$ contribution.

The LHT model which contains only  $(V-A)\otimes (V-A)$ operators belongs to the 
class of models where the connection between $D$ and $K$ physics is particularly
transparent. A detailed analysis in this context has been performed in \cite{Bigi:2009df}. There it has been found that
\begin{itemize}
\item
{The correlation between the $K$ and $D$ systems implies that large NP effects in $K$ and $D$ decays are possible simultaneously.}
\item
{On the other hand simultaneous large NP effects in $D$ and $B$ decays in the LHT model are unlikely. Indeed while either the CP asymmetry $S_{\psi\phi}(B_s)$ in $B_s-\bar B_s$ mixing or $|q/p|$ in the $D^0-\bar D^0$ system can deviate significantly from their SM3 predictions, it is unlikely to observe large deviations from the SM3 in both quantities simultaneously. The improved measurements of $S_{\psi\phi}(B_s)$ at the Tevatron and the LHC in the coming years will therefore have a large impact on the possible size of CP violating effects in $D$ decays within the LHT model.}
\item
{The above observations are consistent with the fact that large simultaneous effects in $K$ and $B$ decays in the LHT are found to be very unlikely \cite{Blanke:2006eb,Blanke:2008ac,Blanke:2009am}.}
\end{itemize}

It is of interest to understand whether this particular pattern of correlations
between $D^0-\bar D^0$ and $K^0-\bar K^0$ mixing can also be found in the SM4.
Analysing the structure of the SM4 contributions we find,
although the SM4 again only involves purely left-handed operators, that NP contributes
in a different manner as compared to the LHT model or the general type of models 
studied in \cite{Blum:2009sk}. The most important difference is the non-unitarity of the $3 \times 3$ submatrix of the CKM matrix in the SM4. Therefore not only the short-distance Wilson coefficients
of the $(V-A)\otimes(V-A)$ operators are modified, but also the CKM elements
in front of the SM3 contributions generally depend on the 4G mixing angles and phases.
From this, it is immediately clear that---in contrast to to the situation assumed in
\cite{Blum:2009sk}---the primary interest is not to use
 $D^0-\bar D^0$ and $K^0-\bar K^0$ mixing observables in order to derive a
lower bound on the relevant NP mass scale. In the contrary, after fixing
reasonable ranges for the $t^\prime$ and $b^\prime$ masses from considerations in
the electroweak sector, 
the flavour observables now generally depend in a rather complicated manner on the 4G mixing
parameters, the short-distance loop functions involving 4G and SM3 quarks, and
long-distance matrix elements.
Focusing on the $D^0-\bar D^0$ sector and using the standard parametrisation
of the 4G mixing matrix together with $\theta_{ij} \ll 1$,
one finds the rather simple approximate expressions
\begin{align}
& \lambda_s^{(D)} \simeq s_{12} \,,
 \qquad 
 \lambda_b^{(D)} \simeq s_{13} \, s_{23} \, e^{-i\delta_{13}} \,,
 \qquad 
 \lambda_{b^\prime}^{(D)} \simeq \sigma_{12} \equiv s_{14} s_{24} e^{i (\delta_{24}-\delta_{14})} \,.
\end{align}
We observe that the new contributions from the $b^\prime$-quark essentially constrain
the magnitude and phase of the parameter combination $\sigma_{12}$. In particular,
we expect sizable contributions to $D^0-\bar D^0$ mixing observables if 
$\lambda_{b^\prime}^{(D)}$ competes with $\lambda_{b}^{(D)}$ appearing in front of 
the short-distance contribution in the SM3, i.e.\ if
\begin{align}
\label{eq:magnitude_constrain}
 s_{14} s_{24} & \gsim  s_{13} \, s_{23} \sim \lambda^5 \,,
\end{align}
where we indicated the scaling with the Wolfenstein parameter $\lambda=s_{12}$.
The analogous expressions for the CKM factors $\lambda_t^{(K)}$ and 
$\lambda_{t^\prime}^{(K)}$, relevant for $K^0-\bar K^0$ mixing, 
can be found  in Section~9.2 of \cite{Buras:2010pi}. 
Here, it is generally found that the expressions depend on all five
new 4G mixing parameters. Consequently, the correlation between 
$D^0-\bar D^0$ and $K^0-\bar K^0$ mixing observables is rather involved.
Some simplification arises if one concentrates on specific scenarios, 
where the 4G mixing angles scale with particular powers of the Wolfenstein parameter $\lambda$, but still
the different loop functions for $t$ and $t^\prime$-quarks prevent us from deriving 
simple analytical expressions as compared to the decoupling scenarios
considered in \cite{Blum:2009sk}.

\section{Numerical Procedure and Results}\label{sec:numerics}
In this section we are going to present the results of our numerical analysis.
First we will discuss the general procedure we have used. Subsequently we will discuss our results for various observables recalling, where necessary, some definitions from our previous papers on $D^0-\bar D^0$ mixing.
For all plots we will use, if not indicated otherwise, the same colour-coding as
in our previous analysis of the SM4 \cite{Buras:2010pi} where we dealt with {mixing and rare decays in the $K$ and $B$ meson systems.}
{This colour coding is given in Table \ref{tab:Bscenarios} and allows to distinguish between sets of SM4 parameters for which the CP asymmetry $S_{\psi\phi}(B_s)$ in the $B_s$ system and the branching ratio $\rm{Br}(B_s\to\mu^+\mu^-)$ assume certain values.} Furthermore, the {\it light blue} 
points correspond to $\rm{Br}(\klpn)>2\cdot 10^{-10}$, while {\it dark 
blue} points indicate $\rm{Br}(\klpn)\le 2\cdot 10^{-10}$. This colour coding will give us some
insight into the correlations between the {different mesonic systems.}
\begin{table}[t!!!pb]
\begin{center}
\begin{tabular}{|c||c|c|c|}
\hline
 		&BS1 (yellow) &BS2 (green)	& BS3 (red) 	\\ \hline\hline
$S_{\psi\phi}(B_s)$	& $0.04\pm 0.01$& $0.04\pm 0.01$ & $ \geq 0.4$ 	\\ \hline
${\rm Br}(B_s\to\mu^+\mu^-)$	& $(2\pm 0.2)\cdot 10^{-9}$ & $(3.2\pm 0.2)\cdot 10^{-9} $   &   $\geq 6\cdot 10^{-9}$ 	\\ \hline
\end{tabular}
\caption{Three scenarios for  $S_{\psi\phi}(B_s)$ and  ${\rm Br}(B_s\to\mu^+\mu^-)$.} \label{tab:Bscenarios}
\end{center}
\end{table}

\subsection{Strategy for Our Phenomenological Analysis}
For our analysis we use the points generated for our previous analysis \cite{Buras:2010pi}. 
For the LD contributions
$\Gamma_{12}^{\rm LD}$ and $M_{12}^{\rm LD}$ we adopt the same procedure as in \cite{Altmannshofer:2010ad}:
We scan flatly over the intervals
\begin{align}
-0.02\,{\rm ps}^{-1}\leq & M_{12}^{\rm LD} \leq 0.02\,{\rm ps}^{-1}\,,\\
-0.04\,{\rm ps}^{-1}\leq & \Gamma_{12}^{\rm LD}\leq 0.04\,{\rm ps}^{-1}\,,
\end{align}
while requiring $x_D,\,y_D$ and $|q/p|$ to be within $2\sigma$ of their experimental values. The latter with their $1\sigma$ ranges are given by \cite{Barberio:2007cr,Barberio:2008fa,Schwartz:2009jv}
\begin{align}\label{eqn:expdata}
x_D  &= \left(0.98^{+0.24}_{-0.26}\right)\%\,, &y_D  &= \left(0.83\pm 0.16\right)\%\,,\nonumber\\
|q/p| &=\left(0.87^{+0.17}_{-0.15}\right)\,, &\varphi &=\left(-8.5^{+7.4}_{-7.0}\right)^\circ\,,\\
\eta_f S_f(D) &=\,\left(-0.248\pm 0.496\right)\%\,,\nonumber
\end{align}
with $\varphi$ being the phase of $q/p$ and the asymmetry $\eta_fS_f(D)$ defined in (\ref{eq:GASYM}).
BaBar recently presented preliminary results for $x_D$ compatible with $x_D=0$. Because of the large relative error 
and the presence of potentially large LD contributions, we will however not elaborate on these new results and use the HFAG averages.
\begin{table}[t!!!pb] 
\begin{center}
\begin{tabular}{|l|l||l|l|}
\hline
parameter & value & parameter & value \\
\hline\hline
$\eta_{cc}$ & $1.51\pm 0.24$\hfill\cite{Herrlich:1993yv}			& $m_{D}$ & $(1.86484\pm0.00017)$GeV\\
$\eta_{tt}$ & $0.5765\pm 0.0065$\hfill\cite{Buras:1990fn}			& $\bar\tau_{D}$ & $(0.4101\pm 0.0015)$ps\\
$\eta_{ct}$ & $0.47\pm 0.04$\hfill\cite{Herrlich:1995hh}			& &\\
$F_D$& $(0.212\pm {0.014})\gev$\hspace{12pt}\cite{Lubicz:2008am}&	 $m_c(m_c)$ & $(1.268\pm 0.009) \gev$\hspace{12pt}\cite{Laiho:2009eu,Allison:2008xk}	 \\
$\hat B_D$& $1.18^{+0.07}_{-0.05}$ & 	$m_b(m_b)$ & $(4.20^{+0.17}_{-0.07}) \gev$\hfill\cite{Barberio:2007cr}	\\
\hline
\end{tabular}
\caption{Values of the input parameters used in our analysis.} \label{tab:parameters}
\end{center}
\end{table}

\boldmath\subsection{The SD Contribution to $D^0-\bar D^0$ Mixing}
\begin{figure}[htbp]
\begin{center}
	\includegraphics[width=0.68\textwidth]{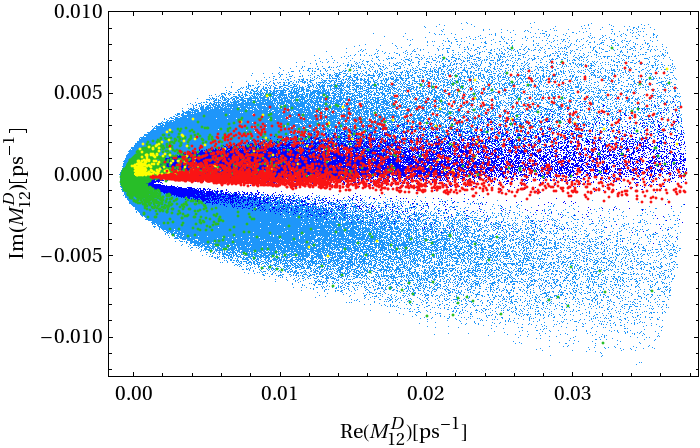}
\caption{The imaginary part as a function of the real part of $M_{12}^D$.\label{fig:m12}}
\end{center}
\end{figure}

\unboldmath
We now turn our attention to the SD contributions which are summarised by the
amplitude $M_{12}^D$ in (\ref{M12D}). In Fig.~\ref{fig:m12} we show
this amplitude in the complex plane.
The striking features of this plot are the following
\renewcommand{\labelenumi}{\roman{enumi})}
\begin{enumerate}
 \item We observe a clustering of the {\textit red} points {(indicating large
     $S_{\psi\phi}(B_s)$ and ${\rm Br}(B_s\to\mu^+\mu^-)$)}
along the $\IM (M_{12}^D)=0$ axis. Furthermore, {for this class of points},
significant positive values of $\text{IM}(M^D_{12})$ are possible, while a rather 
 stringent lower bound, $\IM (M_{12}^D)\gsim -2\cdot 10^{-3}$ is found.
 \item For $\RE (M_{12}^D)>0$, values ranging from very small {up to a factor of four larger} than the {maximal} values of $\IM (M_{12}^D)$ are possible. 
 For $\RE (M_{12}^D)<0$, only very small values are allowed due to the stringent $\eps_K$ constraint.
 \item The largest values for $\IM (M_{12}^D)$ are found in the {\it light blue}
   scenario corresponding to large values of $\rm{Br}(\klpn)$, {which is} governed by CP violation in the K system.
  {We conclude that large CP violating effects in the $D$ system entail large CP violating effects also in the $K\to\pi\nu\bar\nu$ system, while the 
  converse is not necessarily true.}
\end{enumerate}
\renewcommand{\labelenumi}{\arabic{enumi})}
\begin{figure}[htbp]
\begin{center}
\includegraphics[width=0.68\textwidth]{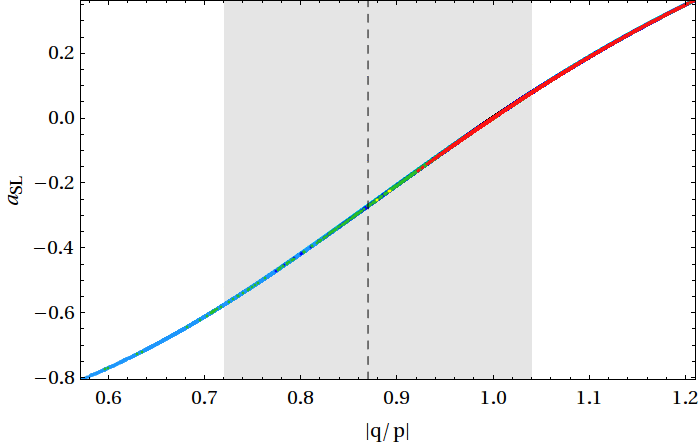}
\caption{The semileptonic CP asymmetry $a_\text{SL}(D)$ as a function of $|q/p|$.\label{fig:qpasl}}
\end{center}
\end{figure}

\subsection{CP Violation in the $D^0-\bar D^0$ System}
In Fig.~\ref{fig:qpasl} we show the semileptonic CP asymmetry $a_\text{SL}(D)$ as a function of $|q/p|$.
{This asymmetry represents CP violation in ${\cal L}(\Delta C=2)$ and is related to $|q/p|$ simply by}
\begin{equation}\label{eq:ASL}
a_\text{SL}(D) \equiv  \frac{\Gamma (D^0(t) \to \ell^-\bar\nu K^{+(*)}) - \Gamma (\bar D^0 \to \ell^+\nu K^{-(*)})}
{\Gamma (D^0(t) \to \ell^-\bar\nu K^{+(*)}) + \Gamma (\bar D^0 \to \ell^+\nu K^{-(*)})}= 
\frac{|q|^4 - |p|^4}{|q|^4 + |p|^4}
\approx 2\left(\left|\frac{q}{p}\right|-1\right)\,.
\end{equation}
In writing the last expression, we have assumed that $|{q}/{p}|-1$ is much smaller
than unity, which is supported by the data listed in (\ref{eqn:expdata}).

We observe---in accordance with Fig.~\ref{fig:m12}---that in the {\it red} scenario, in which CP violation in the $B_s$ system is large, the asymmetry $a_\text{SL}(D)$ in the $D$ system in the allowed range of $|q/p|$ is small but still much larger than in the SM3:
\begin{align}
\left[a_\text{SL}(D)\right]_\text{SM3}&\approx \,1\cdot 10^{-4}\,.
\end{align}
 Significantly larger values are found in the other scenarios in which $S_{\psi\phi}(B_s)$ is SM-like but $\rm{Br}(\klpn)$ might be strongly enhanced.

Of particular interest is the time-dependent CP asymmetry $S_f$ defined by \cite{Bigi:2009df}
\begin{equation}
\frac{\Gamma(D^0(t) \to f) - \Gamma(\bar D^0(t) \to f)}
{\Gamma (D^0(t) \to f) + \Gamma(\bar D^0(t) \to f)} 
\equiv S_{f}(D) \frac{t}{ 2\overline\tau _D} \,,
\label{eq:GASYM}
\end{equation}
which is given by
\begin{equation}
 \eta_f  S_f(D)\simeq - \left[ y_D\left(\left| \frac{q}{p}\right| -\left| \frac{p}{q}\right|   \right)\cos\varphi -
x_D \left(\left| \frac{q}{p}\right| +\left| \frac{p}{q}\right|   \right) \sin\varphi \right] \,,
\label{eq:Sf}
\end{equation}
where $\eta_f=\pm 1$ is the CP parity of the final state $f$.
{The SM3 prediction for $\eta_fS_f(D)$ is}
\begin{align}
\left[\eta_f S_f(D)\right]_\text{SM3}&\approx\,-2 \cdot 10^{-6}\,.
\end{align}
Other useful related formulae can be found in \cite{Bigi:2009df,Altmannshofer:2010ad,Blum:2009sk,Kagan:2009gb}.

In the absence of significant CP phases in the decay amplitudes for $D^0\to f$,
the asymmetry $S_f$ displays a strong correlation  with the semileptonic asymmetry $a_\text{SL}(D)$  \cite{Bigi:2009df},
\begin{equation}
\label{eq:CORR1}
\eta_f S_f(D)\simeq- \frac{x_D^2+y_D^2}{y_D} a_\text{SL}(D)\,.
\end{equation}
A similar correlation is familiar from the $B_s$ system \cite{Ligeti:2006pm,Blanke:2006ig,Grossman:2009mn}. In the presence of significant phases in decay
amplitudes $D^0\to f$, formula (4.23) in \cite{Bigi:2009df} instead of (\ref{eq:Sf}, \ref{eq:CORR1}) applies.

\begin{figure}[htbp]
\includegraphics[width=0.48\textwidth]{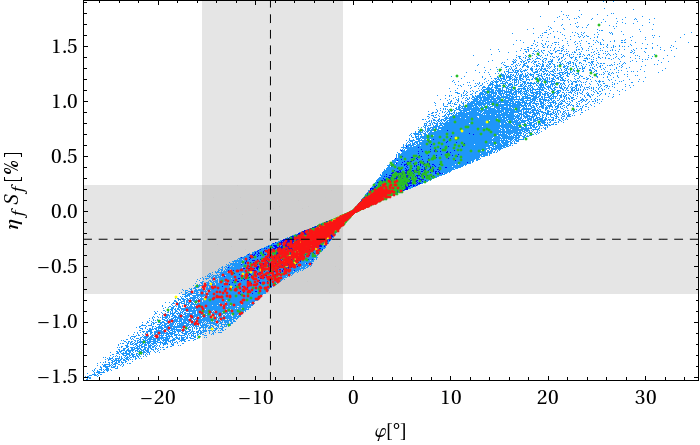}\hspace{0.03\textwidth}
\includegraphics[width=0.48\textwidth]{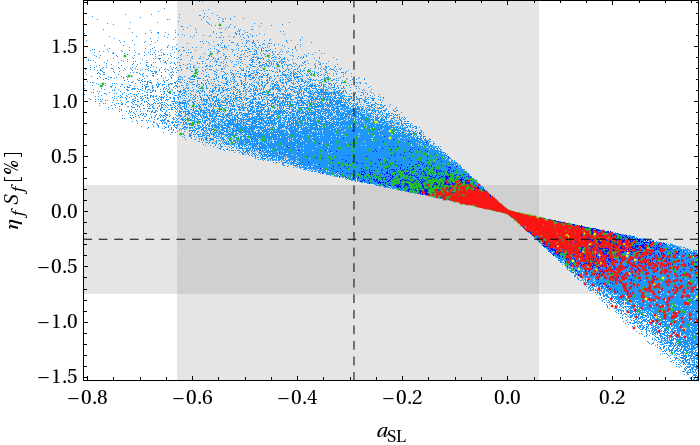}
\caption{Model independent correlations between $\eta_fS_f(D)$ and $\varphi$ (left panel) and $a_\text{SL}$ (right panel).}\label{fig:cpv}
\end{figure}

In Fig.~\ref{fig:cpv} we show $\eta_f S_f(D)$ as a function of the phase $\varphi$ (left panel)
and as a function of $a_\text{SL}(D)$ (right panel).
These plots are familiar from the model-independent analysis in \cite{Altmannshofer:2010ad} and are self-explanatory. In particular in the {\it red} scenario the points cluster around $\eta_f S_f(D) \approx 0$, but still $\eta_f S_f(D)$ can be much larger than found in the SM3. 
{Again extreme values are only found for the {\textit light blue} points corresponding to a large enhancement of ${\rm Br}(\klpn)$.}

\boldmath\subsection{Correlation of $D^0-\bar D^0$ Mixing with the $K$ and $B$ Systems}
\begin{figure}[htbp]
\includegraphics[width=0.48\textwidth]{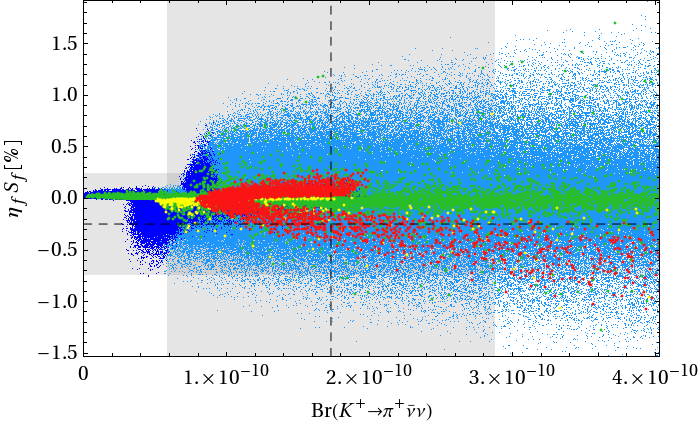}\hspace{0.03\textwidth}
\includegraphics[width=0.48\textwidth]{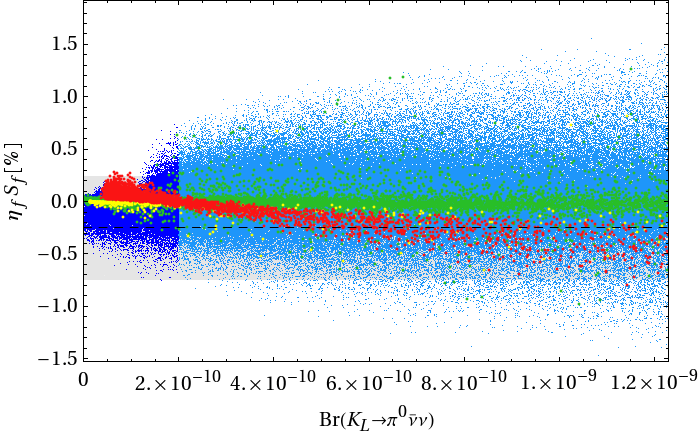}
\caption{$\eta_f S_f(D)$ as a function of ${\rm Br}(\kpn)$ (left panel) and ${\rm Br}(\klpn)$ (right panel).}\label{fig:sfkpn}
\end{figure}

\begin{figure}[htbp]
\includegraphics[width=0.48\textwidth]{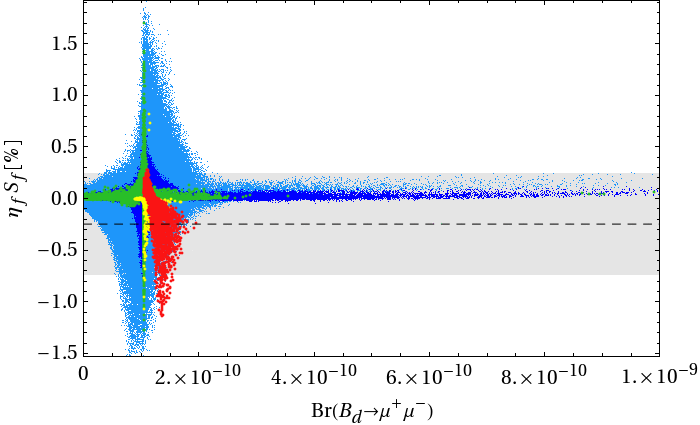}\hspace{0.03\textwidth}
\includegraphics[width=0.48\textwidth]{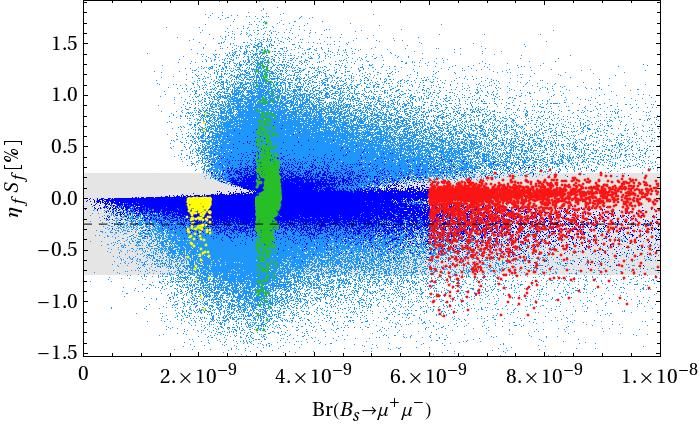}
\caption{$\eta_f S_f(D)$ as a function of ${\rm Br}(B_d\to \mu^+\mu^-)$ (left panel) and ${\rm Br}(B_s\to \mu^+\mu^-)$ (right panel).}\label{fig:sfbq}
\end{figure}

\begin{figure}[htbp]
\begin{center}
	\includegraphics[width=0.68\textwidth]{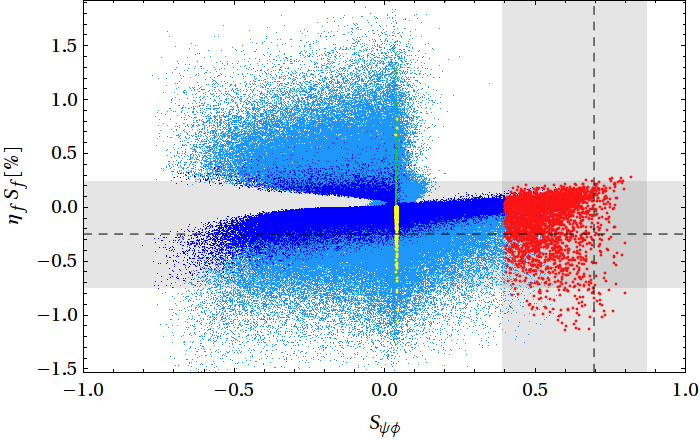}
\caption{$\eta_f S_f(D)$ as a function of $S_{\psi\phi}(B_s)$.\label{fig:sfspsiphi}}
\end{center}
\end{figure}

\begin{figure}[htbp] 
\includegraphics[width=.48\textwidth]{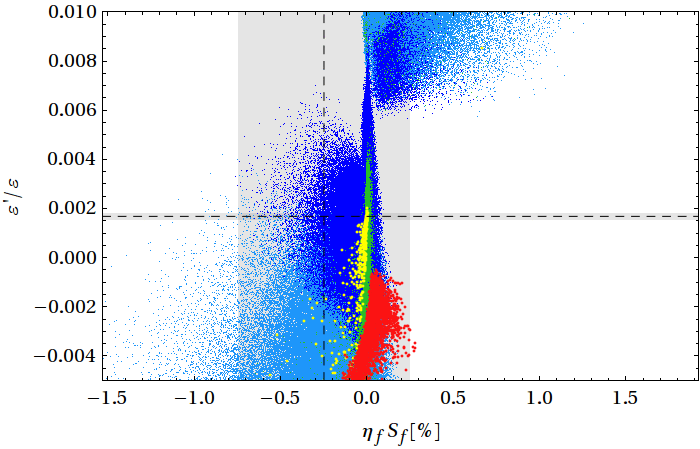}\hspace{.03\textwidth}
\includegraphics[width=.48\textwidth]{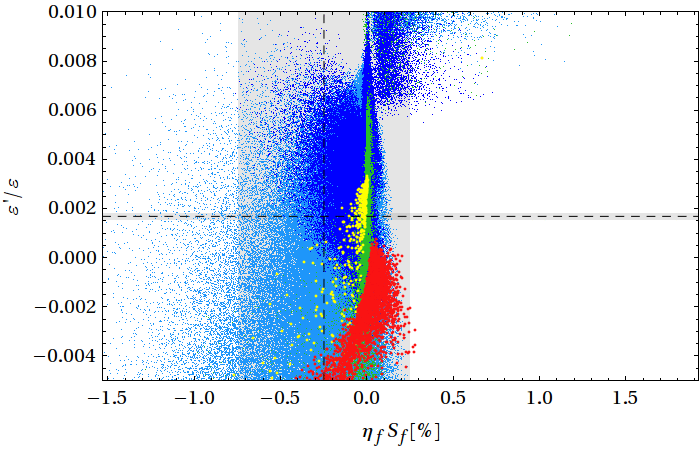}\\
\includegraphics[width=.48\textwidth]{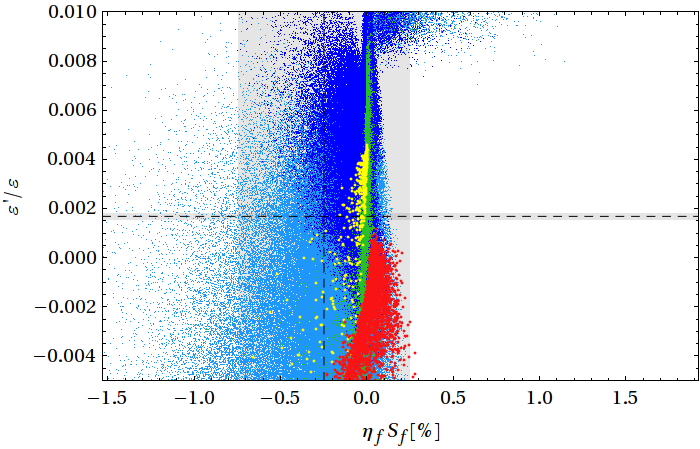}\hspace{.03\textwidth}
\includegraphics[width=.48\textwidth]{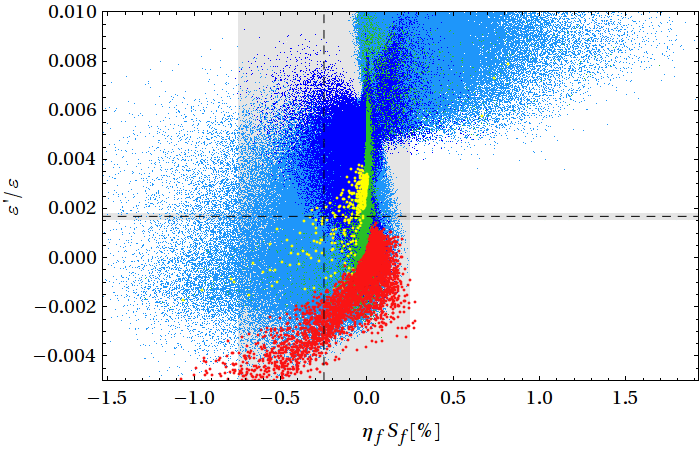}
\caption{$\epe$ as a function of the CP asymmetry $\eta_fS_f(D)$ for four different scenarios of the non-perturbative parameters:
$(R_6,R_8)=(1.0,1.0)$ (upper left panel), $(1.5, 0.8)$ (upper right panel), $(2.0,1.0)$ (lower left panel) and $(1.5,0.5)$ (lower right panel).\label{fig:spsiphiepe}}
\end{figure}

\unboldmath
In order to get some insight into the correlations of the $D$ system with other
meson systems, we show in Figs.~\ref{fig:sfkpn}--\ref{fig:spsiphiepe} a
symphony of plots which depict the dependence of 
$\eta_fS_f(D)$ on a number of prominent observables in the $K$ and $B_{s,d}$ systems:
\begin{itemize}
\item
In the left and right panels of Fig.~\ref{fig:sfkpn} we plot $\eta_fS_f(D)$ as a function of ${\rm Br}(\klpn)$ and 
${\rm Br}(\kpn)$.
We observe that values of $|\eta_f S_f(D)|$ up to $\sim1\%$ are found in the region where ${\rm Br}(\klpn)$ and ${\rm Br}(\kpn)$ are strongly enhanced, but also $\eta_f S_f(D) = 0$ is possible in this region.
\item
In the left and right panels of Fig.~\ref{fig:sfbq} we plot $\eta_fS_f(D)$ as a function of ${\rm Br}(B_d\to\mu^+\mu^-)$ and 
${\rm Br}(B_s\to\mu^+\mu^-)$, respectively.
As expected, for very large values of ${\rm Br}(B_s\to\mu^+\mu^-)$, $\eta_f S_f(D)$ tends to be within the experimental $1\sigma$ range and thus small. 
On the other hand, we find that even for SM-like values of $B_s\to\mu\mu¯$ and $S_{\psi\phi}(B_s)$, maximal values in $\eta_fS_f(D)$ are attainable. Both effects are even 
more pronounced in the case of ${\rm Br}(B_d\to\mu^+\mu^-)$.
{Thus for some values of ${\rm Br}(B_d\to\mu^+\mu^-)$ in conjunction with $\eta_f S_f(D)$ we are able to predict ranges for ${\rm Br}(\klpn)$. The $D$ system in this sense acts as a link between the $K$ and $B$ systems.}

\item
In Fig.~\ref{fig:sfspsiphi} we plot $\eta_f S_f(D)$ as a function of the mixing-induced CP asymmetry $S_{\psi\phi}(B_s)$ in the $B_s$ 
system.
While many of the features in this plot are self-explanatory on the basis of previous plots, the striking result is that for $S_{\psi\phi}(B_s) > 0$ as 
signalled by CDF and D0 data, $\eta_fS_f(D)$ is predicted to be {within its experimental $1\sigma$ range and thus small.}
\item 
{As was pointed out in \cite{Buras:2010pi}, the measured value of $\epe$ practically eliminates large CP violating effects in the $B_s$ system.
In order to study this constraint in the $D$ system, in Fig.~\ref{fig:spsiphiepe} we show  $\epe$ as a function of $\eta_fS_f(D)$ for different values of the non-perturbative parameters $(R_6,R_8)$.
We observe that $\epe$ additionally implies} \textit{negative} or \textit{small positive} values for $\eta_fS_f(D)$. Large negative values 
 for $\eta_fS_f(D)$ however are only possible if also significant CP violation from 4G in the $K\to\pi\nu\bar\nu$ system is present.
\end{itemize}

\FloatBarrier

\boldmath\subsection{Interplay between CP Violation in the $K$, $D$ and $B_s$ Systems\label{sec:seriously}}
\begin{figure}[htbp]
 \begin{center}
  \includegraphics[width=.48\textwidth]{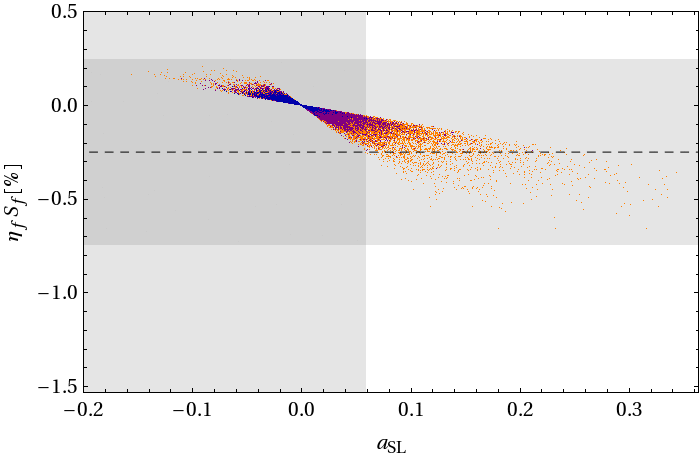}\hspace{.03\textwidth}
\includegraphics[width=.48\textwidth]{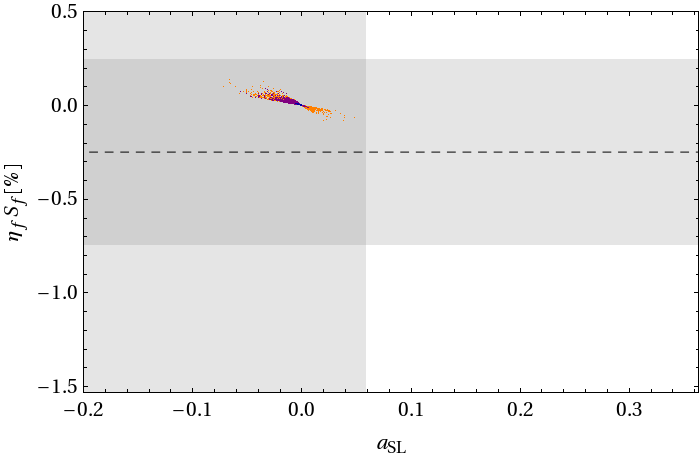}
 \end{center}
\caption{\label{fig:extremeplot}
 Prediction for the CP violating parameters $\eta_f S_f(D)$ vs.~$a_\text{SL}(D)$ in
 the charm system for 4G parameters restricted to fulfil 
 the 2$\sigma$ range for $\varepsilon^\prime/\varepsilon$ and 
$S_{\psi\phi}(B_s) > 0$ (left panel) or $S_{\psi\phi}(B_s)>0.2$ (right panel).
The colour coding is according to Table 5 in \cite{Buras:2010pi} {and indicates the particular values of the hadronic parameters $R_6$ and $R_8$ entering $\epe$.}}
\end{figure}

\unboldmath
Let us, for illustration, consider an extreme situation where we take
the constraints on the 4G parameters from the present experimental and
theoretical situation regarding $\epe$ and $S_{\psi\phi}(B_s)$
at face value. To this end, we restrict our set of allowed 4G parameter values 
to fulfil the 2$\sigma$ range for $\epe$ and either
$S_{\psi\phi}(B_s) > 0$ or $S_{\psi\phi}(B_s)>0.2$. For the remaining points, we plot
again $\eta_f S_f(D)$ as a function of $a_\text{SL}(D)$ and thus obtain a rather 
constrained prediction for the impact of the 4G in the charm system{, as shown in Fig.~\ref{fig:extremeplot}.}
{In particular, we find that the requirement $S_{\psi\phi}(B_s)>0$ in conjunction with the imposition of the constraint from $\epe$ drastically reduces the possible NP effects with respect to those shown in Fig.~\ref{fig:cpv}, but still allows for moderate effects in $a_\text{SL}(D)$ and $\eta_fS_f(D)$, where $\eta_fS_f(D)<[\eta_fS_f(D)]_\text{SM3}$ is slightly favoured. Demanding $S_{\psi\phi}(B_s)>0.2$ on the other hand precludes almost any deviation from the SM3 prediction, regardless of the values of the hadronic parameters $R_6$, $R_8$ entering $\epe$.}
\FloatBarrier

\subsection{Scaling Scenarios}
\begin{figure}[tpbh]
 \begin{center}
{\renewcommand{\arraystretch}{6}
  \begin{tabular}{c c c c c}
 1a: & \parbox[c]{0.4\textwidth}{\includegraphics[width=0.38\textwidth]{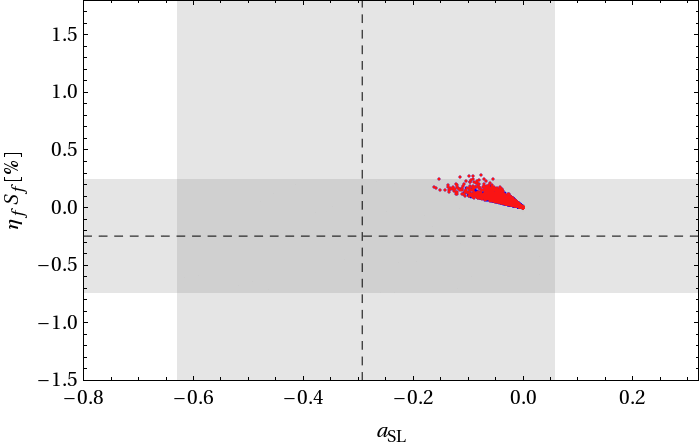}}
&&
 1b: & \parbox[c]{0.4\textwidth}{\includegraphics[width=0.38\textwidth]{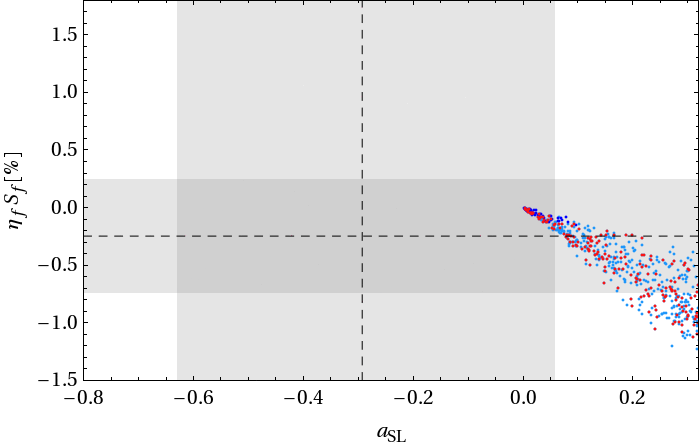}}
\\
 2a: & \parbox[c]{0.4\textwidth}{\includegraphics[width=0.38\textwidth]{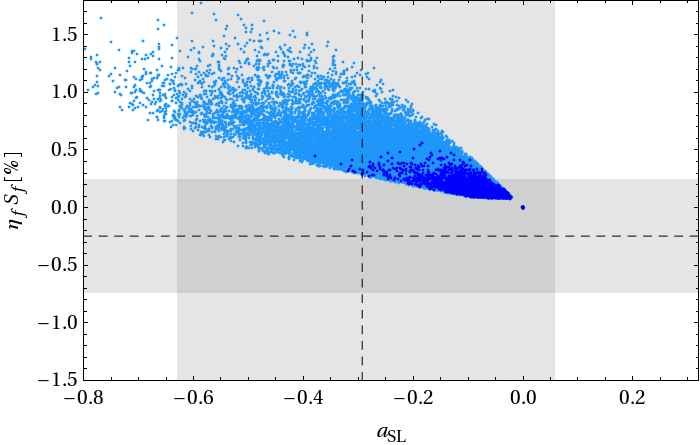}}
&&
 2b: & \parbox[c]{0.4\textwidth}{\includegraphics[width=0.38\textwidth]{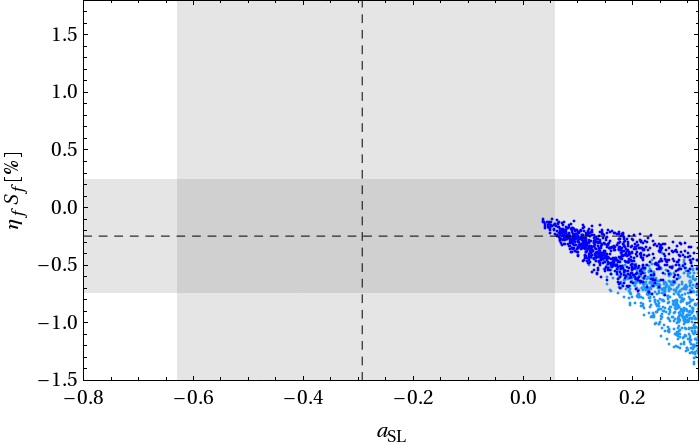}}
\\
 3a: & \parbox[c]{0.4\textwidth}{\includegraphics[width=0.38\textwidth]{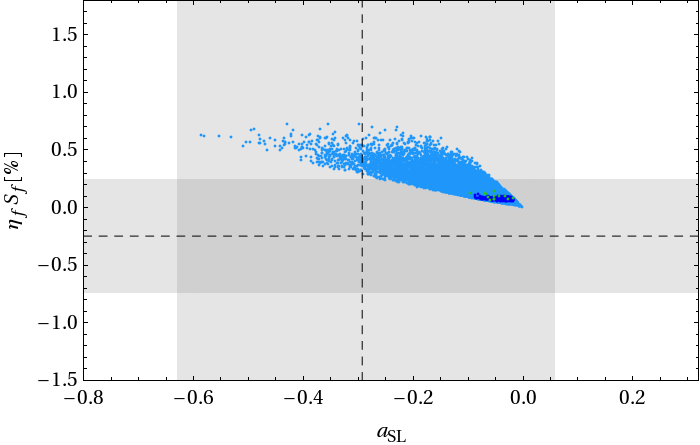}}
&&
 3b: & \parbox[c]{0.4\textwidth}{\includegraphics[width=0.38\textwidth]{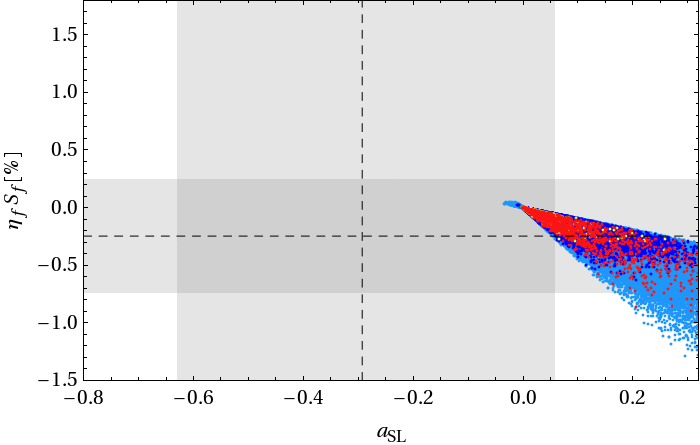}}
\\
 6: & \parbox[c]{0.4\textwidth}{\includegraphics[width=0.38\textwidth]{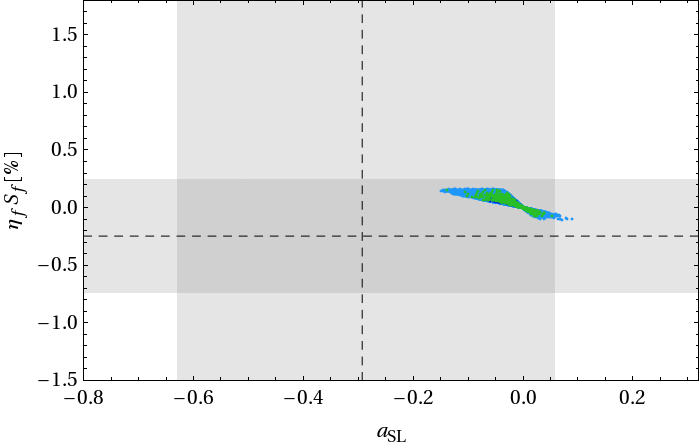}}
&&
 7: & \parbox[c]{0.4\textwidth}{\includegraphics[width=0.38\textwidth]{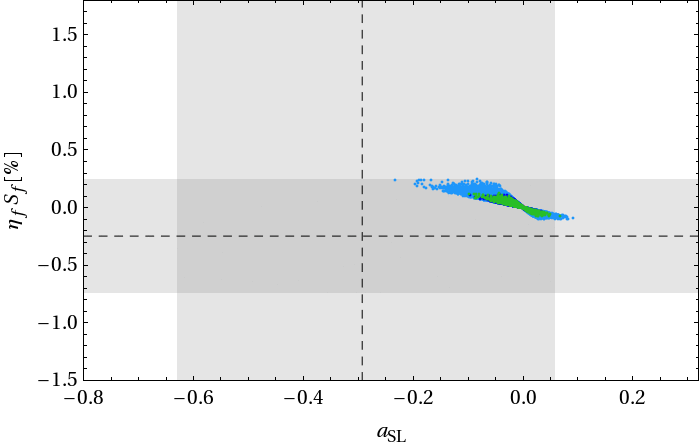}}
  \end{tabular}
}
 \end{center}
\caption{Correlation between $\eta_f S_f(D)$ and 
$a_{\rm SL}(D)$ for
the individual subclasses defined in \cite{Buras:2010pi}.
The colour coding of the points is as explained in Section~\ref{sec:numerics}. \label{fig:DDscenarios}}
\end{figure}

In \cite{Buras:2010pi}, we have shown that it is useful
to characterise the hierarchical structures in the 4G mixing matrix
by dividing the parameter space into subclasses, which are defined by
the scaling of the 4G mixing angles with the Wolfenstein parameter $\lambda$
and the correlation between {the 4G CP phases $\delta_{14}$ and $\delta_{24}$ which is induced by the constraints from precision flavour
observables.} Let us, as an example, investigate the correlation between
the time-dependent CP asymmetry $\eta_f S_f(D)$ and the semileptonic asymmetry
$a_{\rm SL}(D)$ in the $D^0$ decays for
the individual subclasses defined in \cite{Buras:2010pi}, see Fig.~\ref{fig:DDscenarios}.
(We have not shown results for subclasses~4 and 5, as they only lead to tiny
deviations from the SM, since condition (\ref{eq:magnitude_constrain}) is not fulfilled.)
As in the case of $B$-meson and $K$-meson observables, the correlations between
$a_\text{SL}(D)$ and $\eta_fS_f(D)$ are quite different for different
sub-classes, allowing again for a clear separation with respect to
the possible scaling behaviour of the 4G mixing angles.
{Similar plots can be drawn for the other correlations studies above.}

{As a further example we study the 4G parameter points identified in Section \ref{sec:seriously} that are consistent with both the recent experimental data from CDF and D0 implying sizable $S_{\psi\phi}(B_s)$, and the measurement of $\epe$.}
When studying the scaling of the 4G mixing angles and the
correlation between the 4G phases for these points, we find---for
example requiring that $S_{\psi\phi}(B_s)>0.2$---that the
bulk of parameter values has $(\theta_{14},\theta_{24},\theta_{34}) \sim (\lambda^{3},\lambda^2,\lambda)$ and $ -\pi < \delta_{14} \simeq \delta_{24} < 0$.
In terms of the subclasses which we have discussed above, this corresponds
to class~1a (and to a somewhat lesser extent class~3b). A few points with
$(\theta_{14},\theta_{24},\theta_{34}) \sim (\lambda^{4},\lambda^3,\lambda)$
survive, too, but as we already explained above, they correspond to the
uninteresting case where the 4G effects in the charm system are very small.
{Thus, as was the case in our previous analysis \cite{Buras:2010pi}, we find that there is a strong connection between the phenomenological predictions of the SM4 on the one hand and the scaling of the mixing parameters $\theta_{i4}$ in powers of the Wolfenstein parameter $\lambda$ on the other hand.}

\FloatBarrier

\section{Conclusions}\label{sec:conclusion}
In the present paper we have analysed the impact of a 4G of quarks on the
$D^0-\bar D^0$ system taking into account all existing constraints. In addition
we have investigated the correlations of the $D$ system with the prominent
observables in the $K$ and $B_{d,s}$ systems. The main messages from our analysis are as follows:

\begin{itemize}
 \item Large 4G effects in the CP violating observables $a_\text{SL}(D)$ and $\eta_f S_f(D)$ are possible
 and consistent with the available data. The effects are found to be as large as a few times ${\mathcal O}(10^{-1})$ for $a_{\rm SL}(D)$ and
 up to $\sim1.5\%$ for  $\eta_f S_f(D)$, although this seems to be disfavoured by a small $\varphi_{\rm exp}\sim -8^\circ$. Compared to the
 SM3 predictions for these asymmetries, $a_{\rm SL}(D)\sim 10^{-4}$ and $\eta_f S_f(D) \sim 10^{-6}$, even the more moderate enhancements 
{allowed} by the current data are still spectacular.
 \item The large 4G effects in the $K$, $B_d$ and $B_s$ systems found by us in \cite{Buras:2010pi} are still compatible with the current constraints from
 	the charm data.
 \item While simultaneous large 4G effects in the $K$ and $D$ systems are possible, large effects in $B_d$ generally disfavour large NP effects in the $D$ system.
  In the $B_s$ system large CP {\it conserving} effects are found to be possible regardless of the predictions for the $D$ system, while large CP {\it violating} effects can only occur 
  if $\eta_fS_f(D)$ does not deviate from the experimental measurement by much more than $1\sigma$.
{ Conversely, significant enhancement of $S_{\psi\phi}(B_s)$ above the SM3 value will not allow large CP violating effects in the $D$ system with the 4G scenario.}
\item {Additional imposition of the $\epe$ constraint significantly diminishes 4G effects in CP violating observables in the $D$ system.} {This observation is to a large extend independent of the actual values of the hadronic parameters $R_6$ and $R_8$ entering $\epe$ that still suffer from significant theoretical uncertainties.}
\end{itemize}

In the light of these findings we are looking forward to improved data on charm and its correlations with $B$ and $K$ physics measurements to be discovered in the upcoming decade.

\subsection*{Acknowledgements}
{We thank A.~Kronfeld for the illuminating discussion related to the hadronic parameter $\hat B_D$.}
AJB, BD, TF and TH want to thank the Galileo Galilei Institute for Theoretical Physics for their
hospitality. AJB and TF thank the INFN for partial support during the initial steps of this
work. This research was partially supported by the Cluster of Excellence `Origin and Structure of the Universe', the Graduiertenkolleg GRK 1054 of DFG and by the German `Bundesministerium f{\"u}r Bildung und Forschung' under contract 05H09WOE.

%
%
%

\providecommand{\href}[2]{#2}\begingroup\raggedright\endgroup

\end{document}